\documentstyle[aps,floats]{revtex}

\begin{document}
\draft

\title{Gravitational Radiation From Cosmological Turbulence}

\author{Arthur Kosowsky$^{1,2\,\dagger}$, 
Andrew Mack$^{1\,\dagger}$,
and Tinatin Kahniashvili$^{1,3\,\ddagger}$}
\address{$^1$Department of Physics and Astronomy, Rutgers University,
136 Frelinghuysen Road, Piscataway, New Jersey 08854-8019 \\
$^2$School of Natural Sciences, Institute for Advanced Study,
Olden Lane, Princeton, New Jersey 08540 \\
$^3$Center for Plasma Astrophysics, Abastumani Astrophysical Observatory,
A.~Kazbegi Ave.~2a, 380060 Tbilisi, Georgia \\
$^{\dagger}$kosowsky,\,andymack@physics.rutgers.edu, 
$^{\ddagger}$tinatin@amorgos.unige.ch }
\date{April, 2002} 
\maketitle

\begin{abstract}
An injection of energy into the early Universe on a given
characteristic length scale will result in turbulent motions of the
primordial plasma. We calculate the stochastic background of
gravitational radiation arising from a period of cosmological
turbulence, using a simple model of isotropic Kolmogoroff turbulence
produced in a cosmological phase transition. We also derive the
gravitational radiation generated by magnetic fields arising from a
dynamo operating during the period of turbulence.  The resulting
gravitational radiation background has a maximum amplitude comparable
to the radiation background from the collision of bubbles in a
first-order phase transition, but at a lower frequency, while the
radiation from the induced magnetic fields is always subdominant to
that from the turbulence itself. We briefly discuss the detectability of
such a signal.

\end{abstract}


\section{Introduction}

Gravitational radiation is likely the only direct source of
information about the Universe at very early times. Electromagnetic
radiation has propagated freely only since the epoch of recombination
at a redshift $z\simeq 1000$; any radiation produced at earlier times
was quickly thermalized by Compton scattering from free electrons in
the primordial plasma. Neutrinos probe to somewhat earlier epochs
since they were in thermal equilibrium only until the Universe was
around one second old, but detection prospects for the cosmic neutrino
background are nil. In contrast, gravitational radiation was in
thermal equilibrium only at temperatures approaching the Planck energy
when the Universe had an age of around the Planck time. Furthermore,
gravitational radiation, unlike electromagnetic radiation, propagates
virtually unimpeded throughout the entire history of the Universe.
These properties make gravitational radiation a powerful probe of the
very early Universe, in principle. The difficulty is of course the
extremely small amplitude of the propagating metric perturbations.

The most cosmologically interesting gravitational radiation sources
are stochastic backgrounds produced by some event in the early
evolution of the Universe. One widely discussed example is the
background of tensor metric perturbations produced by quantum
fluctuations during inflation \cite{inflation}. However, the amplitude
of temperature fluctuations in the cosmic microwave background likely
limits the amplitude of an inflationary gravitational wave background
to be undetectably small on scales amenable to direct detection (i.e.\
laboratory to solar system scales).  Another possibility is a
significant background from the evolution of topological defects such
as cosmic strings \cite{strings}. Current measurements of the
microwave background and the large-scale distribution of galaxies
rule out defects as the sole structure formation mechanism, although
it is conceivable that some small fraction of the microwave background
fluctuations arise from defects. In this case as well, direct
detection of the gravitational radiation from defects appears
improbable.

The most promising source of a detectable cosmological background of
stochastic gravitational waves is a phase transition in the early
Universe \cite{wit84,hog86}. A first-order phase transition proceeds
via the random nucleation of bubbles of the new phase, which
subsequently expand and merge, converting the old phase to the new
phase. The coherent motion of the bubble walls, which contain a
significant fraction of the free energy associated with the phase
transition, can produce copious gravitational radiation
\cite{tw90,ktw92a,ktw92b}. The radiation spectrum generically peaks at
a comoving wavelength corresponding to the Hubble length at the time
of the phase transition times the bubble wall velocity in units of the
speed of light.  Remarkably, the horizon scale at the electroweak
phase transition falls into the frequency band of the proposed 
Laser Interferometer Space Antenna (LISA)
space-based laser interferometric gravitational radiation detector
\cite{LISA}, and a reasonably strong electroweak phase transition
(although much stronger than in the standard model) would be
detectable with currently planned gravitational wave experiments
\cite{kkt94}.

Besides the bubble wall motions in a phase transition, a related
source of gravitational radiation is the subsequent turbulent motion
of the plasma following the phase transition. Dimensional analysis
suggests that turbulence might contribute a gravitational radiation
background comparable to or larger than that from bubble wall motions
\cite{kkt94,hog00}. In the absence of bubble shape instabilities, the
bubbles of the low-temperature phase will expand spherically until
encountering other expanding bubbles. After the bubbles collide, a
region of complex, turbulent plasma motions will result since large
amounts of energy are being injected on a particular characteristic
length scale. As the phase transition completes, the bubble wall
motions sourcing the turbulence cease to be effective, and the
turbulence damps away with a characteristic damping time scale
depending on the plasma viscosity. If the bubbles are unstable to
distortions of their shapes, then the expansion of the non-spherical
bubbles can also create additional turbulence. If the turbulence is
strong, with velocities some non-negligible fraction of the speed of
light, significant gravitational radiation can be generated during the
interval between the initial bubble collisions and the damping of the
turbulence after the completion of the phase transition.

In this paper, we quantify these claims by computing the gravitational
radiation resulting from an idealized turbulent source. We assume that
a source of turbulence exists for some specified length of time,
injecting energy on a particular length scale at a particular
redshift. We model the resulting turbulence as having a Kolmogoroff
energy spectrum. Details of the turbulence model and discussion of
the validity of various assumptions are presented in Sec.~II.  We
then compute the generated gravitational waves using the turbulent
plasma motions as a source to the wave equation (Sec.~III); the
results are then converted to present-day amplitudes and energy
densities as functions of frequency.  Section IV derives the additional
gravitational radiation generated by turbulence-induced magnetic
fields, showing that the peak amplitude from this source will be 
far smaller than the peak amplitude from the
turbulence itself, though at a higher frequency.  
In Sec.~V, we apply the results to a generic
model of first-order phase transitions, including a brief review of
hydrodynamic bubble evolution.  Section VI discusses the detectability
of the resulting backgrounds with planned and envisioned experiments.
Throughout the paper we employ natural units with $c=\hbar=k_B=1$.

A substantial literature on cosmological turbulence appeared
three decades ago, when turbulent vorticity was considered as
a mechanism for initiating galaxy formation \cite{protogalaxies}. 
While this particular idea
soon fell out of favor due to inconsistency with the microwave
background isotropy \cite{anile76} and nucleosynthesis \cite{bar77}, 
some formal aspects of these treatments
are relevant for this work; see, e.g., \cite{oc68,nar74,ko74},
which develop phenomenological descriptions of cosmological
turbulence similar in spirit to that presented in this paper.
The hydrodynamic equations in an expanding Universe were derived
through a transformation of the nonexpanding case in \cite{ds76},
a special case of a more general theorem \cite{sb98}.

We emphasize that our results are independent of the nature of the
turbulence source. While first-order phase transitions are the only
obvious source of strong turbulence in the early Universe, the
calculations presented here are equally applicable to any other
potential source of turbulence (see, e.g., \cite{dg02}).

\section{Model Isotropic Turbulence}

The theory of turbulence was originally formulated over sixty years
ago \cite{richardson,kol41}.  But the complexity of turbulent motion
makes any analysis beyond basic scaling considerations and dimensional
analysis intractable. Model isotropic turbulence is experimentally
tested via wind tunnel measurements on scales small compared to the
size of the tunnel, and the concepts of a cascade of kinetic energy
from large to small scales and the role of viscosity are well
established.  But classical turbulence analysis is done for
non-relativistic fluid velocities and incompressible fluids. Here we
need to model turbulence in a radiation-dominated plasma, potentially
with moderately relativistic fluid velocities and complications like
shock formation.  While the theory of turbulence in highly
relativistic plasmas is not well understood, we will simply extend the
nonrelativistic results in the naive manner with the understanding
that some corrections might apply.

Consider an event in the early Universe, presumably a first-order
phase transition, which converts an energy density $\kappa\rho_{\rm
vac}$ into kinetic energy of the primordial plasma in some
characteristic time scale $\tau_{\rm stir}$ 
on some characteristic source length
scale $L_S$. Here $\rho_{\rm vac}$ is the total free energy density
liberated and $\kappa$ is an efficiency factor which accounts for the
fraction of the available energy which goes directly into kinetic, as
opposed to thermal, energy. The length scale $L_S$ must be connected
to the Hubble length $H_*^{-1}\simeq m_{\rm Pl}/T_*^2$, which is the
only cosmological length scale at early times; we write $L_S\equiv
\gamma H_*^{-1}$. Here $T_*$ is the temperature of the Universe when the
event takes place and $m_{\rm Pl}$ is the Planck mass. Under suitable
conditions discussed below, a turbulent cascade will develop in which
energy will be transferred from larger to smaller scales as eddies of
progressively smaller sizes are formed from larger ones. The cascade
stops at a damping scale $L_D$ when the fluid kinematic viscosity
$\nu$ diffuses the turbulent velocities at the same rate as they are
replenished from larger scales. We assume that for scales $L$ in the
range $L_D < L < L_S$ (the inertial range), the turbulence is
homogeneous and isotropic.  We also must know the enthalpy density
$w=\rho + p$ of the (nonturbulent) plasma, which appears in the
stress-energy tensor.  In our simplified model, any turbulent source
in the early Universe is determined completely by the physical
quantities $\rho_{\rm vac}$, $\kappa$, $\tau_{\rm stir}$, 
$L_S$, $T_*$, $w$, and
$\nu$. These quantities in turn determine $L_D$, the damping scale,
and $\tau$, the total duration of the turbulence.
 Note that a given cosmological model determines $w$ and $\nu$
from the temperature $T_*$.  We also define the wave numbers $k_S =
2\pi/L_S$ and $k_D = 2\pi/L_D$ corresponding to the largest and
smallest turbulence scales.

The turbulent energy in the cascade
is characterized by the stationary Kolmogoroff spectrum
\begin{equation}
E(k)\equiv {1\over w}
{d\rho_{\rm turb}\over dk} = C_k {\bar\varepsilon}^{2/3} k^{-5/3},
\label{KolEk}
\end{equation}
where $\rho_{\rm turb}$ is the kinetic energy density of the turbulent
motions.  The Kolmogoroff constant $C_k$ is of order unity and
$\bar\varepsilon$ is the energy dissipation rate per
unit enthalpy given by \cite{my75}
\begin{equation}
{\bar\varepsilon}= 2\nu \int_{k_S}^{k_D} dk\,k^2 E(k)
\label{dissipation}
\end{equation}
where $\nu$ is the kinematic viscosity of the plasma.  This spectrum
holds for a constant rate of energy flow from larger to smaller
scales; the amplitude is fixed by the rate of energy dissipation.  For
a non-relativistic plasma, the enthalpy density $w$ is just the mass
density of the plasma, while for temperatures large compared to the
masses of particles in the plasma or for any radiation-dominated
plasma, $w$ is $4/3$ times the thermal energy density of the plasma.
Combining the above two equations and solving for the energy
dissipation rate gives
\begin{equation}
{\bar\varepsilon} \simeq {27\over 8}k_D^4 \nu^3
\label{epsilonbar}
\end{equation}
assuming $C_k=1$ and $k_S \ll k_D$. However, $E(k)$ is
not yet determined since we do not know the wave number $k_D$
corresponding to the smallest-scale turbulent motions.

Before completing the specification of $E(k)$ in terms of the physical
variables defining the phase transition, consider the time scales
involved in the turbulence. Assume that the only peculiar velocities
present are (relativistic) turbulent velocities with spatial distribution 
${\bf u}({\bf x})$; we employ the Fourier convention
\begin{equation}
{\bf u}({\bf k}) = {1\over V}\int d{\bf x}\, e^{i{\bf k}\cdot{\bf x}}
{\bf u}({\bf x}).
\label{u_k}
\end{equation}
We retain the fiducial volume factor $V$ to insure consistent
dimensions for all quantities; all physical results will
be independent of $V$.
A statistically isotropic and homogeneous velocity field of an 
incompressible fluid has the two-point correlation function
\begin{equation}
\left\langle u_i({\bf k}) u^*_j({\bf k}') \right\rangle
= {(2\pi)^3\over V} P_{ij}({\bf\hat k}) P(k) 
\delta({\bf k} - {\bf k}'),
\label{uuexpectation}
\end{equation}
where 
\begin{equation}
P_{ij}({\bf\hat k}) \equiv \delta_{ij} - {\hat k}_i {\hat k}_j
\label{Pij}
\end{equation}
is a projector onto the transverse plane:
\begin{equation}
P_{ij}P_{jk}=P_{ik},\qquad\qquad P_{ij}{\hat k}_j = 0.
\label{pijproperties}
\end{equation}
The angular brackets in Eq.~(\ref{uuexpectation})
mean a statistical average when the
velocities are considered as random variables
(see Ref.~\cite{my75}, Volume 1, for a detailed discussion).
If the fluid is compressible, a second arbitrary function
appears in the correlation function, proportional to
${\hat k}_i {\hat k}_j$, describing longitudinal motions.
A specific model for isotropic turbulence consists of specifying
the function $P(k)$; we assume the power spectrum is a power law,
$P(k) = A k^n$,
where the normalization $A$ and the spectral index
$n$ can be deduced from the Kolmogoroff spectrum.
The mean square velocity of the fluid at any point in space is
given by
\begin{equation}
\left\langle {\bf u}^2({\bf x})\right\rangle 
= {V\over(2\pi)^3}\int d{\bf k}\, 2P(k)
= {V\over\pi^2} \int_{k_S}^{k_D} dk\, k^2P(k).
\label{meansqvel}
\end{equation}
But this quantity is just the kinetic energy density per unit 
enthalpy density of the fluid; thus we derive the connection
\begin{equation}
E(k) = {V\over \pi^2} k^2 P(k).
\end{equation}
For the case of a Kolmogoroff spectrum, Eq.~(\ref{KolEk}) implies that
\begin{equation}
P(k) \simeq {1\over V}\pi^2{\bar\varepsilon}^{2/3} k^{-11/3}.
\label{PkKol}
\end{equation}

We are interested in the characteristic eddy velocity on a given scale
$L$. From the slope of the Kolmogoroff spectrum and
Eq.~(\ref{meansqvel}), it follows that the total turbulent velocity at
a given point is dominated by the eddy velocity on the largest
scale. We can thus estimate the characteristic eddy velocity on the
scale $L$ by cutting off the integral in Eq.~(\ref{meansqvel}) at a
wave number $k_L = 2\pi/L$ corresponding to that scale:
\begin{eqnarray}
u_L &\simeq& \left[ \int_{k_L}^{k_D} dk\,E(k)\right]^{1/2}\cr
    &=& \left(3\over 2\right)^{1/2}(2\pi)^{-1/3}
             ({\bar\varepsilon}L)^{1/3}.
\label{uL}
\end{eqnarray}
We can also estimate an eddy turnover time scale (known as
the circulation time) on a length scale
$L$ as the ratio of $L$ to the physical velocity  $v_L = u_L/(1+u_L^2)^{1/2}$. 
We argue below that the physical velocity will be approximately
bounded by the sound speed of the fluid; for a radiation-dominated
plasma, this condition is $v_L \leq 1/\sqrt{3}$. Making the simple
approximation that $v_L=u_L$ until the sound speed is reached, after
which time $v_L$ is the sound speed, the circulation time is
\begin{equation}
\tau_L \simeq L/v_L \simeq \cases{
{3\over 2}{\bar\varepsilon}^{-1/3} L^{2/3},
& $L \leq 3^{3/2}(8\bar\varepsilon)^{-1}$;\cr
L\sqrt{3},& otherwise.\cr}
\label{tauL}
\end{equation}

Now the remaining undetermined quantity in the turbulence spectrum,
$k_D$, can be fixed via energy considerations. Two different
cases must be considered separately, depending on whether the
duration of the turbulent source $\tau_{\rm stir}$ 
is long or short compared to
the eddy turnover time scale $\tau_S$ on the characteristic
length scale of the source $L_S$. First consider the simpler case
where $\tau_{\rm stir} \gg \tau_S$. Fully developed turbulence is
established in a time on the order of $\tau_S$, so this case
gives approximately a stationary source lasting for a time 
$\tau = \tau_{\rm stir}$. 
To keep the turbulence stationary, the energy dissipation rate
must equal the mean input power of the source:
\begin{equation}
\bar\varepsilon = {\kappa\rho_{\rm vac}\over w\tau_{\rm stir}}.
\label{epsbarvalue1}
\end{equation}
This expression immediately determines the amplitude of the
Kolmogoroff spectrum, Eq.~(\ref{KolEk}), and comparing with
Eq.~(\ref{epsilonbar}) gives
\begin{equation}
k_D \simeq
\left({8\kappa\rho_{\rm vac}\over 27\nu^3\tau_{\rm stir} w}\right)^{1/4}.
\label{kDvalue1}
\end{equation}
Thus the turbulent gravitational wave source is completely determined
for this case. The circulation time scale on the scale of the source
is approximated by combining Eqs.~(\ref{tauL}) and (\ref{epsbarvalue1})
to give
\begin{equation}
\tau_S \simeq {3\over 2} \left(L_S^2\tau_{\rm stir} w\over\kappa\rho_{\rm vac}
\right)^{1/3},
\label{tauS}
\end{equation}
so the condition for this case to be valid becomes
\begin{equation}
\tau_{\rm stir} \gg L_S \left(w\over\kappa\rho_{\rm vac}\right)^{1/2}.
\label{condition1}
\end{equation}
Finally, the Reynolds number for this turbulence is given by
\begin{equation}
{\rm Re} = \left(k_D\over k_S\right)^{4/3} 
\simeq {2\over 3}\left({1\over 2\pi}\right)^{4/3}
\left({\kappa\rho_{\rm vac} L_S^4\over \nu^3\tau_{\rm stir} w}\right)^{1/3}.
\label{reynolds}
\end{equation}
The critical Reynolds number for the onset of stationary turbulence
is around 2000. Early Universe phase transitions will generally
have Reynolds numbers exceeding this value.

The alternate case, for $\tau_{\rm stir} \ll \tau_S$, 
is more subtle. Here, an
impulsive force is imparted to the plasma, resulting in a total
kinetic energy density equal to the total free energy density of the
phase transition times the efficiency factor $\kappa$, coherent on the
length scale $L_S$. The efficiency factor depends on the mechanical
details of the stirring process and will be a function of mean
input power $\rho_{\rm vac}/\tau_{\rm stir}$.
A cascade of kinetic energy to smaller scales will occur, but
stationary, isotropic turbulence will never develop because the plasma
is not continually being stirred by the source. We can estimate the
time for which significant kinetic energy on a given scale lasts.  On
the largest scale $L_S$, the kinetic energy will last for a time set
by the dissipation time scale, approximately equal to the eddy
turnover time $\tau_S$. As in fully developed turbulence, this kinetic
energy will cascade to smaller scales.  The eddies on the largest
scale will act as a source for eddies on a slightly smaller scale $L$
for a time $\tau_S$. On the smaller scale, we assume the plasma has no
kinetic energy at the moment of the impulsive force but rather
acquires kinetic energy only from the cascade.  The smaller-scale
eddies are spun up in a time corresponding to the circulation time on
the smaller scale $\tau_L$; these eddies will last until the
large-scale source becomes ineffectual and then will dissipate also on
the circulation time scale $\tau_L$.  So by this argument, the eddies
on a smaller scale $L$ will exist for the same total amount of time as
the eddies on the largest scale $L_S$, although their establishment
and dissipation will be displaced to a slightly later time compared
with the largest-scale eddies. The same reasoning can then be applied
to eddies at successively smaller scales, with the following
conclusion: on any given scale between $L_S$ and $L_D$, eddies will
exist for a total time $\tau_S$. The only assumption required for this
conclusion is that the time scale for establishing eddies on a given
scale via the cascade from larger scales is the same as the time scale
for dissipating the same eddies via the cascade to smaller scales.

The time displacements of the time intervals for the existence of
eddies on different scales are essentially irrelevant for the
generation of gravitational radiation, leading only to some relative
phase shift between the gravitational radiation at two different
frequencies. Therefore, for the purposes of modelling a gravitational
wave source, we assume the plasma motion consists of kinetic energy
simultaneously on all scales within the inertial range, lasting for a
total time $\tau = \tau_S$, the circulation time on the scale of the
turbulence source, with a kinetic energy density spectrum given by the
Kolmogoroff spectrum, Eq.~(\ref{KolEk}). To normalize the spectrum, we
simply treat the total free energy density as being injected
continually over the time $\tau_S$ rather than as an impulse. Now this
case looks just like the previous one, except that $\tau_{\rm stir}$ must be
replaced by $\tau_S$ in Eqs.~(\ref{kDvalue1}) and (\ref{tauS}):
\begin{equation}
k_D \simeq\left({8\kappa\rho_{\rm vac}\over 
27\nu^3\tau_S w}\right)^{1/4},\qquad\qquad
\tau_S = L_S \left({3\over 2}\right)^{3/2}
\left({w\over\kappa\rho_{\rm vac}}\right)^{1/2}.
\label{kDvalue2}
\end{equation}
Combining the two cases gives the simple expressions
\begin{equation}
k_D \simeq\left({8\kappa\rho_{\rm vac}\over 
27\nu^3\tau w}\right)^{1/4},\qquad\qquad
\tau_S \simeq {3\over 2} \left(L_S^2\tau w\over\kappa\rho_{\rm vac}
\right)^{1/3},
\label{kDvalueboth}
\end{equation}
valid for either case, where 
\begin{equation}
\tau = {\rm max}(\tau_S,\,\tau_{\rm stir}).
\label{taudef}
\end{equation}
The eventual expression for the gravitational wave
amplitude is only very weakly dependent upon $\tau$, 
so the distinction between the two cases is largely
unimportant for our results.

When computing the gravitational wave signal, we will encounter
unequal time velocity correlators of the form
\begin{equation}
\left\langle u_i({\bf k},t) u^*_j({\bf k}',t') \right\rangle
\equiv {(2\pi)^3\over V} P_{ij}({\bf\hat k}) F(k,t-t') 
\delta({\bf k} - {\bf k}')
\label{unequaltime}
\end{equation} 
[cf.\ Eq.~(\ref{uuexpectation})]. The 
dependence of the function $F$ only upon the time difference
$t-t'$ follows from the assumption that the
turbulence can be treated as stationary, with $F(k,0) = P(k)$.  No
general form is known for $F$. However, general physical
considerations imply that $F$ must be a decreasing function of $t-t'$,
and we assume that the decay of $F$ should have a characteristic time scale on
the order of the circulation time on the scale $L=2\pi/k$. We
actually will only need to guarantee that $F$ goes to zero
no faster than the light-crossing time of $L$, which is guaranteed
by causality.

We have sidestepped the issue of relativistic versus nonrelativistic
turbulence. The Kolmogoroff model of turbulence phenomenology has only
been formulated and tested for turbulence with nonrelativistic
velocities in plasmas with nonrelativistic equations of state.  No
general model exists for the opposite situation of a relativistic
plasma with relativistic velocities. The plasma in our case will also
be compressible, contrary to the basic assumption above. For a large
enough input of energy, plasma velocities may be driven past the sound
speed, leading to shock formation. We conservatively assume that the
sound velocity represents an upper limit to the turbulent plasma
velocity, because shocks will result in significant thermal
dissipation. Note that to the extent that shock fronts retain kinetic
energy, our ultimate gravitational wave background will be increased
relative to the estimates made here, in the case of highly
relativistic fluid velocities.

To summarize, this model for cosmological turbulence requires (i)
$\kappa\rho_{\rm vac}$, the energy density converted to turbulent
motion where $\rho_{\rm vac}$ is a characteristic energy density and
$\kappa$ is an efficiency factor; (ii) $L_S$, the characteristic length
scale of the source producing the turbulence (the ``stirring scale'');
(iii) $\tau_{\rm stir}$, the duration of the source producing the
turbulence; (iv) $T_*$, the temperature of the Universe at the onset of
the turbulence, which in turn determines $w$, the enthalpy density,
and $\nu$, the kinematic viscosity of the plasma.  The assumption of
stationary homogeneous and isotropic Kolmogoroff turbulence then
specifies in terms of these quantities (i) the normalization of the
turbulence power spectrum, (ii) the length scale $L_D$ at which the
turbulence is dissipated by viscosity, and (iii) the circulation time
for any particular turbulent length scale between $L_S$ and $L_D$.

We have neglected the expansion of the Universe in this description
of turbulence. If the duration of the turbulence $\tau$ is longer
than the Hubble time $H^{-1}$, then the expansion will produce
additional damping of the turbulence as the energy density is
redshifted. Furthermore, if the circulation time on the stirring
scale $\tau_S$ is comparable to or longer than the Hubble time,
the expansion damping may inhibit the establishment of a
turbulent cascade. Particular cases should be checked
individually, but in general, if a phase transition is strong
enough to drive turbulence producing an interestingly large
gravitational radiation amplitude, it will last for a time
short compared to the Hubble time and expansion damping will
be negligible. This claim can be quantified using the
expressions derived in Sec.~\ref{sec:pt} below.

\section{Gravitational Radiation from Turbulent Plasma}
\label{sec:gw}

\subsection{General Considerations}

The source of gravitational radiation is the transverse and
traceless piece of the stress-energy tensor of a given system.
For turbulent plasma, the relevant stress-energy tensor is given by
\begin{equation}
T_{ij}({\bf x}) = w u_i({\bf x}) u_j({\bf x}).
\label{Tijdef}
\end{equation}
The above expression drops the diagonal (trace) component of the
stress-energy because it cannot source any gravitational radiation.
To simplify the problem, we assume (conservatively) that the enthalpy
density $w$ remains
constant throughout space, while the variation of the velocity vector
describes the turbulent motions of the plasma. If this assumption does
not hold, the resulting gravitational wave amplitude will increase.
In Fourier space, the stress-energy is then given by the convolution
\begin{equation}
T_{ij}({\bf k},t) \simeq {V\over (2\pi)^3}w\int d{\bf q}\, u_i({\bf q},t)
u_j({\bf k-q},t).
\label{TijFT}
\end{equation}

Gravitational radiation is produced by the transverse and
traceless piece of the stress-energy tensor. Given an
arbitrary stress-energy tensor in Fourier space, $T_{ij}({\bf k},t)$,
the portion sourcing gravitational radiation can be obtained
by applying a projection tensor (see, e.g., \cite{mtw73}):
\begin{equation}
\Pi_{ij} = \left(P_{il} P_{jm} - {1\over 2}P_{ij}P_{lm}\right) T_{lm}.
\label{Piij}
\end{equation}
Once the source is specified, the gravitational wave metric
perturbations $h_{ij}$ obey the wave equation
\begin{equation}
{d^2h_{ij}\over d\eta^2} + {2\over a}{da\over d\eta} {dh_{ij}\over d\eta} 
+ {\tilde k}^2 h_{ij} = 8\pi G a^2 \Pi_{ij}
\label{hij_evolution}
\end{equation}
where $\eta$ is conformal time, ${\tilde k}$ is the comoving wave number, and
$a$ is the scale factor of the Universe. 
Note that we have defined the tensor metric perturbation as
$\delta g_{ij}\equiv 2h_{ij}$.  For relevant phase
transitions, the duration of the source will be short compared
to the Hubble time, which means the expansion of the Universe
can be neglected during the generation of the waves. We can
thus drop the expansion drag term in Eq.~(\ref{hij_evolution})
and change variables to physical time and physical wave number,
obtaining the simple oscillator equation
\begin{equation}
{\ddot h}_{ij}({\bf k},t) + k^2 h_{ij}({\bf k},t) 
= 8\pi G \Pi_{ij}({\bf k},t),
\label{hij_flat}
\end{equation}
where dots denote derivatives with respect to $t$. From this
point on, all wave numbers will refer to physical, not
comoving, quantities. 

The source considered here turns on at a specific time $t_*$
and we assume no gravitational radiation exists prior
to this time. The initial conditions for Eq.~(\ref{hij_flat})
are simply $h_{ij}({\bf k},t_*) = {\dot h}_{ij}({\bf k},t_*) = 0$. 
In the Euclidean space approximation we have made, the radiation
generated cannot depend on the particular value of $t_*$, so 
for convenience we set $t_*=0$ in this section. Of course,
once the results are translated back into expanding spacetime,
the time $t_*$ of the phase transition fixes the energy
and length scale associated with the phase transition.
The Green function for the homogeneous equation is simply
\begin{equation}
G(t,t') = \cases{
0, & $0<t<t',$\cr
{1\over k}\sin\left[k(t-t')\right], & $0<t'<t,$\cr
}
\label{greenfunction}
\end{equation}
with $G={\dot G} =0$ at $t=0$. The general solution
for the wave amplitude is then
\begin{equation}
h_{ij}({\bf k},t) = {8\pi G\over k} \int_{0}^\tau \Theta(t-t') 
\sin\left[k(t-t')\right] \Pi_{ij}({\bf k},t') dt'
\label{generalsolution}
\end{equation}
where $\Theta$ is a step function.

\subsection{Time Averaging Technique}

Since turbulence is a stochastic process, we cannot
compute the exact gravitational waveforms. Our goal
is to compute the average power spectrum or characteristic
amplitude of the waves. We are concerned here only
with the power spectrum, so consider the quantity
\begin{eqnarray}
\left\langle h_{ij}({\bf k},t)h^*_{ij}({\bf k'},t)\right\rangle
&=& {(8\pi G)^2\over V} \delta({\bf k}-{\bf k'})\cr
&&\times\left\langle{1\over k^2}\int_0^\tau dt_1 \int_0^\tau dt_2\,
\Theta(t-t_1)\Theta(t-t_2)
\sin\left[k(t-t_1)\right]\sin\left[k(t-t_2)\right]
\Pi_{ij}({\bf k},t_1)\Pi^*_{ij}({\bf k},t_2)\right\rangle.
\label{hh1}
\end{eqnarray}
The delta-function factor is guaranteed by statistical
isotropy of the gravitational waves; we have written this
dependence out explicitly and then changed all factors
of ${\bf k'}$ to ${\bf k}$ within the
angular brackets. To make further progress, we need a practical way
to deal with the averaging process. We are assuming a
stationary, homogeneous and isotropic source, so we
make the simple assumption that the statistical
average can be estimated by either a time or
space average. To evaluate Eq.~(\ref{hh1}) we
use a time average, since all of the time dependence
is in the Green functions and not in the source terms.
Then we have
\begin{eqnarray}
\left\langle h_{ij}({\bf k},t)h^*_{ij}({\bf k'},t)\right\rangle
&=& \delta({\bf k} - {\bf k'})
{(8\pi G)^2 \over Vk^2}\int_0^\tau dt_1 \int_0^\tau dt_2\,
\Pi_{ij}({\bf k},t_1)\Pi^*_{ij}({\bf k},t_2)\cr
&&\qquad \times {1\over T}\int_s^{s+T}dt\,
\Theta(t-t_1)\Theta(t-t_2)
\sin\left[k(t-t_1)\right]\sin\left[k(t-t_2)\right],
\label{hh2}
\end{eqnarray}
where $s$ is some arbitrary time when the source is active,
and $T$ is an interval of time long enough for the average
to be approximated by the time average. In practice, this will
be some time on the order of a few circulation times on a given
scale. As $t_1$ or $t_2$ approaches $\tau$, it will not
be possible to choose $T$ large enough for a rigorously
valid average, but this will not appreciably affect our
estimates since we are considering only statistical averages
for the source terms: the time integration is a convenient
device for approximating the effect of this averaging,
and the averaging itself becomes only a rough approximation
for durations shorter than the circulation time on a given scale.
Since we are assuming a stationary source, Eq.~(\ref{hh2})
must be independent of the chosen value of $s$. We
choose an $s$ which eliminates the step functions from
the integral, keeping in mind the above discussion.

The integral over $t$ is now elementary:
\begin{equation}
{1\over T}\int_s^{s+T}dt
\sin\left[k(t-t_1)\right]\sin\left[k(t-t_2)\right]
= {1\over 2}\cos\left[k(t_2-t_1)\right]
- {1\over 2Tk}\sin(Tk) \cos\left[k(2s + T - t_1 - t_2)\right].
\label{t_int}
\end{equation}
We neglect the second term with respect to the first since
$Tk \gg 1$: $k^{-1}$ will be on the order of the light
crossing time for a given scale, while $T$ will be
at least as long as the circulation time on the given scale
$k^{-1}$, so the comparison will be valid on all scales except
for possibly the largest, where at least the simple inequality
$Tk > 1$ will hold. Since the terms are both oscillatory,
the comparison really only applies to the size of the prefactors,
but this is sufficient for our purpose. Now substituting
Eq.~(\ref{t_int}) into Eq.~(\ref{hh2}) and making the
substitution $y=t_2-t_1$ gives
\begin{equation}
\left\langle h_{ij}({\bf k},\tau)h^*_{ij}({\bf k'},\tau)\right\rangle
\simeq \delta({\bf k} - {\bf k'})
{(8\pi G)^2\over 2Vk^2}\int_{-t_1}^{\tau-t_1} dy \,
\cos(ky) \int_0^\tau dt_1\, 
\Pi_{ij}({\bf k},t_1)\Pi^*_{ij}({\bf k},t_1+y).
\label{hh3}
\end{equation}
We now use the $t_1$ integral as an estimator for the
statistical average of the sources, giving
\begin{equation}
\left\langle h_{ij}({\bf k},\tau)h^*_{ij}({\bf k'},\tau)\right\rangle
\simeq {(8\pi G)^2\tau\over 2kk'}\int_0^\tau 
dy \cos(ky) 
\left\langle\Pi_{ij}({\bf k},t_1)\Pi^*_{ij}({\bf k'},t_1+y)\right\rangle,
\label{hh4}
\end{equation}
where the delta-function has been reabsorbed into the statistical
average. Note that the average on the right side is independent
of $t_1$ since the source is assumed to be stationary.

We now have an expression involving the average source correlation
at different times, integrated against an oscillating function.
Note that the total value is proportional to $\tau$, the duration
of the source, as it should be for an incoherent source.
To make further progress, we require a more explicit form for
the source average. 

\subsection{Evaluation of the Source Average}

We have expressions for averages of the fluid velocities
in the turbulent source; we need to connect these with the
particular average required in Eq.~(\ref{hh4}). 
Writing out the projectors in Eq.~(\ref{Piij}) gives
\begin{eqnarray}
\left\langle\Pi_{ij}({\bf k},t)\Pi^*_{ij}({\bf k'},t+y)\right\rangle
&=& \bigl[P_{ia}({\bf\hat k})P_{jb}({\bf\hat k}) 
- {1\over 2}P_{ij}({\bf\hat k})P_{ab}({\bf\hat k})\bigr]
\bigl[P_{ic}({\bf\hat k'})P_{jd}({\bf\hat k'}) 
- {1\over 2}P_{ij}({\bf\hat k'})P_{cd}({\bf\hat k'})\bigr]\cr
&&\qquad\qquad\qquad\times
\left\langle T_{ab}({\bf k},t) T^*_{cd}({\bf k'},t+y)\right\rangle.
\label{pipi1}
\end{eqnarray}
We need to evaluate the expectation value of the stress
tensor product. Equation (\ref{TijFT}) shows that this
product will involve the expectation value of
four velocity vectors evaluated at two different
times. No general solution is known for such expectation
values for turbulent flow. The simplest (and most conservative)
assumption is that the correlation function factors into
products of pairs of velocities, as for a Gaussian field.
Then Wick's theorem applies and we have
\begin{eqnarray}
&&\left\langle T_{ab}({\bf k},t) T^*_{cd}({\bf k'},t+y)\right\rangle
= {V^2 \over (2\pi)^6} w^2\int d{\bf q}\,d{\bf s}
\Bigl[\left\langle u_a({\bf q},t)u^*_b({\bf q}-{\bf k},t)\right\rangle
\left\langle u_c(-{\bf s},t+y)u^*_d({\bf k'}-{\bf s},t+y)\right\rangle\cr
&& \qquad + \left\langle u_a({\bf q},t)u^*_c({\bf s},t+y)\right\rangle
\left\langle u_b({\bf k}-{\bf q},t)u^*_d({\bf k'}-{\bf s},t+y)\right\rangle
+ \left\langle u_a({\bf q},t)u^*_d({\bf k'}-{\bf s},t+y)\right\rangle
\left\langle u_b({\bf k}-{\bf q},t)u^*_c({\bf s},t+y)\right\rangle \Bigr].
\label{tt1}
\end{eqnarray}
This expression can be simplified using the correlation functions
in Eqs.~(\ref{uuexpectation}) and (\ref{unequaltime}), giving
\begin{equation}
\left\langle T_{ab}({\bf k},t) T^*_{cd}({\bf k'},t+y)\right\rangle
= w^2 \delta({\bf k}-{\bf k'}) \int d{\bf q}
\left[P_{ac}({\bf\hat q})P_{bd}(\widehat{{\bf k}-{\bf q}})
+P_{ad}({\bf\hat q})P_{bc}(\widehat{{\bf k}-{\bf q}})\right]
F(q,y) F(|{\bf k}-{\bf q}|,y).
\label{tt2}
\end{equation}
The first of the three terms in Eq.~(\ref{tt1}) does not
contribute, since it is nonzero only for the constant offset 
mode with ${\bf k}={\bf k'}=0$. After substituting the explicit
form for the projectors, Eq.~(\ref{Pij}), setting 
${\bf k}={\bf k'}$ from the delta function, and simplifying
the contractions, we obtain
\begin{equation}
\left\langle\Pi_{ij}({\bf k},t)\Pi^*_{ij}({\bf k'},t+y)\right\rangle
= w^2\delta({\bf k}-{\bf k'}) \int d{\bf q}\, F(q,y) F(|{\bf k}-{\bf q}|,y)
(1+\gamma^2)(1+\beta^2),
\label{pipi2}
\end{equation}
where we have defined the auxiliary quantities 
$\gamma = \hat{\bf k}\cdot \hat{\bf q}$ and
$\beta = \hat{\bf k}\cdot \widehat{{\bf k}-{\bf q}}$. 

Substituting this simple form for the unequal time source
correlation into Eq.~(\ref{hh4}) gives
\begin{equation}
\left\langle h_{ij}({\bf k},\tau)h^*_{ij}({\bf k'},\tau)\right\rangle
= {(8\pi G)^2\tau w^2\over 2k^2}\delta({\bf k}-{\bf k'})
\int d{\bf q}\,  (1+\gamma^2)(1+\beta^2)
\int_0^\tau dy \cos(ky)F(q,y) F(|{\bf k}-{\bf q}|,y).
\label{hh5}
\end{equation}
Now $F(k,0)=P(k)$ so we make the further assumption
that $F$ can be separated as
\begin{equation}
F(k,y) = P(k) D(yk^{2/3});
\label{separability}
\end{equation}
that is, we have assumed a universal form for the time decay for all
$k$ values, with the time argument of $F$ scaling with the circulation
time on the length scale $2\pi/k$, and $D$ is some monotonically
decreasing function of its argument. This is likely a reasonable
assumption for fully developed turbulence. On the other hand, we are
only concerned with the time dependence to the extent that it is
integrated against the oscillatory function $\cos(kt)$ in
Eq.~(\ref{hh5}). Since $F$ or $D$ is everywhere positive, the integral
itself is oscillatory.  If our crude turbulent model were exact, the
induced power spectrum of gravitational waves would exhibit
oscillations. But this is an artifact of the assumption that the
turbulence begins and ends at precisely defined times. For the present
task of estimating characteristic amplitudes for a realistic
turbulence source, we instead approximate the time integral by its
root-mean-square value. The $\cos(ky)$ term will always oscillate on a
time scale shorter than the characteristic time for $D(yk^{2/3})$, as
seen from a simple comparison of the circulation time to the light
crossing time for a given scale $L$. Thus regardless of the particular
time dependence of $D$, we approximate
\begin{equation}
\int_0^\tau dy \cos(ky)F(q,y) F(|{\bf k}-{\bf q}|,y)
\simeq \int_0^\tau dy \cos(ky)P(q) P(|{\bf k}-{\bf q}|)
\simeq {\sqrt{2}\over 2k}P(q) P(|{\bf k}-{\bf q}|).
\label{cosintegral}
\end{equation}
This approximation replaces the time-dependent function $D$ by the
constant $D(0)$. Actually $D$ will decrease with time. This will {\it
increase} the mean value of the integral unless the characteristic
time scale for the decrease of $D$ is less than $k^{-1}$, which we
have argued will never be obtained, so the approximation in
Eq.~(\ref{cosintegral}) is actually a conservative one.

Substituting this result into Eq.~(\ref{hh5}) and replacing
$\gamma^2$ and $\beta^2$ by their average values of $1/2$
over the integral gives the simple approximate form
\begin{equation}
\left\langle h_{ij}({\bf k},\tau)h^*_{ij}({\bf k'},\tau)\right\rangle
\simeq {9\sqrt{2}(8\pi G)^2\tau w^2\over 16k^3}\delta({\bf k}-{\bf k'})
\int d{\bf q}\, P(q) P(|{\bf k}-{\bf q}|).
\label{hh6}
\end{equation}
We now have an expression which can be evaluated for the
particular turbulent power spectrum to give the final
expression for the power in gravitational radiation in
terms of the turbulence parameters.

\subsection{The Power Spectrum}

For a power law power spectrum, the remaining integral in
Eq.~(\ref{hh6}) is elementary. Using the general form
$P(k)=Ak^n$,
\begin{eqnarray}
\int d{\bf q}\, P(q) P(|{\bf k}-{\bf q}|)
&=& 2\pi A^2\int_{k_S}^{k_D} dq\,q^{n+2} \int_{-1}^1 d\gamma\, 
(k^2+q^2-2kq\gamma)^{n/2}\cr
&=& 4\pi A^2\left[{k^{2n+3}n\over (n+3)(2n+3)} +{k_D^{2n+3}\over 2n+3}
-{k^n k_S^{n+3}\over n+3}\right]. 
\label{PPintegral}
\end{eqnarray}
For the specific case of Kolmogoroff turbulence, the power
law is $n=-11/3$; then
the last term in Eq.~(\ref{PPintegral}) is dominant. Keeping
only this term and inserting Eq.~(\ref{PkKol}) for the power
spectrum gives
\begin{equation}
\left\langle h_{ij}({\bf k},\tau)h^*_{ij}({\bf k'},\tau)\right\rangle
\simeq {216\pi^7\sqrt{2}G^2\over V^2}\tau w^2
{\bar\varepsilon}^{4/3} k^{-20/3}k_S^{-2/3}\delta({\bf k}-{\bf k'}).
\label{hh7}
\end{equation}

To make contact with measurable quantities, we evaluate the
real-space correlation function
\begin{eqnarray}
\left\langle h_{ij}({\bf x},\tau)h_{ij}({\bf x},\tau)\right\rangle
&=& {V^2\over (2\pi)^6}\int d{\bf k}\,d{\bf k'}\, 
e^{i({\bf k'}-{\bf k})\cdot {\bf x}}
\left\langle h_{ij}({\bf k},\tau)h^*_{ij}({\bf k'},\tau)\right\rangle\cr
&\simeq& 
{27\sqrt{2}\pi^2\over 2} G^2\tau w^2
{\bar\varepsilon}^{4/3} k_S^{-2/3} \int_{k_S}^{k_D} dk\, k^{-14/3}.
\label{hh8}
\end{eqnarray}
Now we need to convert this expression to one involving the
gravitational wave frequency $f$. The frequency is determined
by the scale of time variation corresponding to the spatial
Fourier mode $k$, the circulation time $\tau_L$ given
in Eq.~(\ref{tauL}). Writing $f=\tau_L^{-1}$ and changing
variables in the $k$ integral gives
\begin{equation}
\left\langle h_{ij}({\bf x},\tau)h_{ij}({\bf x},\tau)\right\rangle
\simeq {2^{2/3}\over 3^{5/2}\pi^{7/3}} {\bar\varepsilon}^{7/2}
G^2\tau w^2 f_S^{-1} \int_{f_S}^{f_D} df\, f^{-13/2};
\label{hh9}
\end{equation}
the numerical prefactor is about 0.007. We define the
characteristic gravitational wave amplitude $h_c(f)$
per unit logarithmic frequency interval
(following Maggiore \cite{mag00}) via 
\begin{equation}
\left\langle h_{ij}({\bf x},\tau)h_{ij}({\bf x},\tau)\right\rangle
\equiv \frac{1}{2}\int_0^\infty {df\over f} h_c^2(f).
\label{hcdef}
\end{equation}
Note that Eq.~(\ref{hcdef}) is smaller than the corresponding expression
in Ref.~\cite{mag00} by a factor of 4 since the tensor metric perturbation in
Ref.~\cite{mag00} is defined as $\delta g_{ij}\equiv h_{ij}$ whereas ours
is  $\delta g_{ij}\equiv 2h_{ij}$ [see comments after
Eq.~(\ref{hij_evolution})].  Comparing with
Eq.~(\ref{hh9}) gives
\begin{equation}
h_c(f) = 0.12 G {\bar\varepsilon}^{7/4}
w\tau^{1/2}f_S^{-1/2}f^{-11/4}
\label{hc_pt}
\end{equation}
for frequencies between $f_S$ and $f_D$, with the frequency
at the stirring scale
\begin{equation}
f_S \simeq {2\over 3}{\bar\varepsilon}^{1/3}L_S^{-2/3}.
\label{fS}
\end{equation}

\subsection{Relic Gravitational Radiation}

The above expressions apply to the waves generated at the
time of the phase transition. We then stretch the
waves with the expansion of the Universe: the frequency
and amplitude are both inversely proportional to the scale factor. 
The latter follows from the fact that the total energy density
in gravitational radiation scales like $a^{-4}$ with the expansion,
and the energy density is proportional to $\langle \dot h \dot h\rangle$. 
For turbulence generated at a time when the temperature
of the Universe was $T_*$, the ratio of the scale factor
then to the scale factor now is
\begin{equation}
{a_*\over a_0} = 8.0\times 10^{-16}\left(100\over g_*\right)^{1/3}
\left(100\,{\rm GeV}\over T_*\right)
\label{scalefactor}
\end{equation}
where $g_*$ is the number of relativistic degrees of freedom
at the temperature $T_*$. The Hubble parameter at this
time is
\begin{equation}
H_*^2 = {8\pi G\over 3}\rho_{\rm rad} 
= {8\pi^3 g_* T_*^4\over 90 m_{\rm Pl}^2} 
\label{hubbleparam}
\end{equation}
with $m_{\rm Pl}$ the Planck mass. This gives
the relation
\begin{equation}
{\tilde f} = 1.65\times 10^{-5}\,{\rm Hz}\,\left(f_*\over H_*\right)
\left({T_*\over 100\,{\rm GeV}}\right)\left(g_*\over 100\right)^{1/6}
\label{tildef}
\end{equation}
where $f_*$ is a radiation frequency at the cosmic temperature $T_*$
and $\tilde f$ is the corresponding frequency of the radiation today.
Scaling Eqs.~(\ref{hc_pt}) and (\ref{fS}) by the expansion of
the Universe and substituting $w=4\rho_{\rm rad}/3$ 
and ${\bar\varepsilon} = \kappa\rho_{\rm vac}/(w\tau)$
gives 
\begin{equation}
h_c({\tilde f}) = 5.6\times 10^{-17}
\left(\kappa\rho_{\rm vac}\over w\right)^{2/3}
\left(\tau\over H_*^{-1}\right)^{-1/6}
\left(L_S\over H_*^{-1}\right)^{13/6}
\left(100\,{\rm GeV}\over T_*\right)
\left(100\over g_*\right)^{1/3}
\left({\tilde f}\over {\tilde f}_S\right)^{-11/4},
\label{scaledh}
\end{equation}
for the characteristic amplitude,
which holds for ${\tilde f} > {\tilde f}_S$, and
\begin{equation}
{\tilde f}_S = 1.1\times 10^{-5}\,{\rm Hz}\,
\left(\kappa\rho_{\rm vac}\over w\right)^{1/3}
\left(\tau\over H_*^{-1}\right)^{-1/3}
\left(L_S\over H_*^{-1}\right)^{-2/3}
\left(T_*\over 100\,{\rm GeV}\right)
\left(g_*\over 100\right)^{1/6}.
\label{scaledf}
\end{equation}
Equations (\ref{scaledh}) and (\ref{scaledf}) are our fundamental
results. Converting to the characteristic energy density
in gravitational radiation via the relation
\begin{equation}
h_c({\tilde f}) = 1.3\times 10^{-18} \left({\rm Hz}\over {\tilde f}\right)
\sqrt{\Omega_{\rm GW}({\tilde f})h^2},
\label{omega_gw_def}
\end{equation}
where $h$ is the current Hubble parameter in units of 100 km/s Mpc$^{-1}$
and $\Omega_{\rm GW}({\tilde f})$ is the energy density in gravitational waves
per logarithmic frequency interval
in units of the current critical density, gives
\begin{equation}
\Omega_{\rm GW}({\tilde f})h^2 = 2.2\times 10^{-7}
\left(\kappa\rho_{\rm vac}\over w\right)^2
\left(\tau\over H_*^{-1}\right)^{-1}
\left(L_S\over H_*^{-1}\right)^3
\left(g_*\over 100\right)^{-1/3}
\left({\tilde f}\over{\tilde f}_S\right)^{-7/2}.
\label{omega_gw}
\end{equation}

\section{Gravitational Radiation from Induced Magnetic Fields}
\label{sec:magnetic}

In addition to the turbulent motions, gravitational radiation also may
be generated by magnetic fields arising from a turbulent dynamo
mechanism: generically, the turbulence will exponentially amplify any
seed magnetic fields until the field strength saturates at equipartition
with the turbulent kinetic energy. The characteristic $e$-folding
time scale on a give length scale $L$ will be simply the circulation
time $\tau_L$. The mechanism of seed field
generation is not clear, but seed fields might naturally arise during
a phase transition due to bubble wall instabilities combined
with surface charge densities on the bubble walls and 
magnetohydrodynamic amplification \cite{soj97}.  Once a magnetic field
is generated, the high conductivity of the primordial plasma will keep
the field frozen in.

It is reasonable to suspect that such a field may give a significant
background of gravitational radiation: since the magnetic field has a
nonzero stress, it will provide a coherent source term in
Eq.~(\ref{hij_evolution}). Such a magnetic field will act as a
gravitational radiation source from the time of the phase transition
until the field is damped (or until matter-radiation equality,
if the field lasts that long), rather than just during the brief
period of turbulence. The following calculation, however, shows that
induced magnetic fields produce a maximum characteristic
amplitude of gravitational radiation which is always much smaller than
the maximum amplitude from the turbulence which generated
them. The magnetic field gravitational radiation peaks at a 
much higher frequency, though, and can have a larger amplitude than the
turbulence-induced gravitational radiation at that frequency.
As in the previous section, quantities below are physical,
except for comoving quantities denoted with a tilde.

\subsection{General Magnetic Field Considerations}

First, we assume the turbulence-induced magnetic fields are generated
almost instantaneously during the time of the phase transition.  To a
good approximation, the turbulence-induced magnetic fields are
generated within a tiny fraction of the Hubble time $H^{-1}_*$, thus
we can normalize the magnetic field power spectrum at the time of the
phase transition.  Second, we assume the turbulence-induced magnetic
fields are saturated at an equipartition value up to a physical scale
$L_B$ at the time of the phase transition.  We will leave the ratio
between the magnetic field physical saturation scale to the turbulence
stirring scale, i.e.\ $L_B/L_S$, as a free parameter in our final
expressions.  This will make the comparison with the previous
turbulence results easier.  Generally, we expect $L_B/L_S$ to be on
the order of 0.003: the turbulence circulation time scales like
$L^{2/3}$, so $(L_B/L_S)^{2/3}$ gives the ratio of
$e$-foldings of the magnetic field on the scales $L_S$ and
$L_B$. Conservatively estimating that the turbulence lasts for a
single circulation time on the largest scale, a range of $L_B/L_S$
between 0.003 and 0.0017 gives a range of magnetic field amplification
factors between $10^{20}$ and $10^{30}$. The exponential amplification
makes this estimate robust: making the seed fields smaller by a factor
of $10^{10}$ only reduces $L_B$ by a modest fraction.  Third, we
assume the turbulence-induced magnetic fields are just frozen into the
plasma and retain the form of the spectrum until they are damped away
by neutrino viscosity. Damping of magneto-hydrodynamic (MHD) modes
by neutrino viscosity is
most efficient before and around nucleosynthesis $(T\sim 0.1\,{\rm
MeV})$.  At the time of neutrino decoupling $(T\sim1\,{\rm MeV})$, the
neutrino physical mean free path
$(l_{\nu\,\text{dec}}\approx10^{11}\,{\rm cm})$ and the Hubble length
$(H^{-1}_{\nu\,\text{dec}}\approx5\times10^{10}\,{\rm cm})$ are
comparable, hence all the subhorizon magnetic perturbations generated
during the electroweak phase transition will be damped away by the
time of nucleosynthesis (see, e.g., \cite{sb98,jko98}).
We do not consider any kind of inverse-cascade mechanism that will
transfer small-scale magnetic fields to larger scales.  Invoking an
inverse cascade will spread the magnetic energy to scales larger than
$L_B$ and reduce the overall gravity wave amplitude.  This will also
push the gravitational radiation frequencies to smaller values than
those obtained below.

A statistically homogeneous and isotropic stochastic magnetic field has a
two-point correlation function given by Eq.~(\ref{uuexpectation})
with a power spectrum we denote $P_B(k)$. We assume that the 
turbulence-induced magnetic field exists on scales between the
saturation scale $L_B$ and the turbulence damping scale $L_D$. The
mean-square value of the magnetic field is
[see, e.g., Eq.~(2.7) of Ref.~\cite{mack02}]
\begin{equation}
B^2=\frac{V}{\pi}\int^{k_D}_{k_B}dk\, k^2P_B(k).
\label{eq:B-turb-B-mean-square}
\end{equation}
We now normalize $P_B(k)$ using the fact that the turbulence-induced magnetic
field energy density is half of the turbulent kinetic energy density
\begin{equation}
\frac{1}{2}wu^2_B=\frac{B^2}{8\pi}.
\label{eq:B-turb-energy-normalize}
\end{equation}
Using Eqs.~(\ref{meansqvel}) and
(\ref{eq:B-turb-B-mean-square}), we obtain
\begin{equation}
P_B(k)=4\pi wP(k) = \frac{4\pi^3w\bar{\varepsilon}^{2/3}k^{-11/3}}{V}
\label{eq:B-turb-power-spectrum}
\end{equation}
using Eq.~(\ref{PkKol}). 

In Fourier space, the turbulence-induced magnetic stress-energy tensor is
given by the convolution of the magnetic field [see Eq.~(2.9) of
Ref.~\cite{mack02}]:
\begin{equation}
T^{(B)}_{ij}({\mathbf k},t_*)=\frac{V}{(2\pi)^3}\frac{1}{4\pi}
\int d{\mathbf q}\,\left[B_i({\mathbf q},t_*)B_j({\mathbf k-q},t_*)
-\frac{1}{2}\delta_{ij}B_l({\mathbf q},t_*)B_l({\mathbf k-q},t_*)\right],
\label{eq:B-turb-stress}
\end{equation}
where the explicit $t_*$ dependence is to remind ourselves of the
assumption that the turbulence-induced magnetic fields are generated almost
instantaneously during the time of the phase transition.  In addition,
we have neglected the induced electric field due to the fact that the early
Universe is highly conductive. The source for gravitational
radiation is given by the transverse-traceless projection of this
stress tensor, Eq.~(\ref{Piij}). 

In the absence of any inverse-cascade mechanism, magnetic fields are just
frozen into the plasma and evolve by simply redshifting with the Universe's
expansion until they are damped away by neutrino viscosity.  Therefore, 
magnetic fields act on a longer time scale than the turbulent fluid velocities.
To facilitate the computation, we introduce a comoving
quantity $\Pi^{(B)}_{ij}(\tilde{{\mathbf k}})$ corresponding to
$\Pi^{(B)}_{ij}({\mathbf k},t_*)$ via
\begin{equation}
\Pi^{(B)}_{ij}(\tilde{{\mathbf k}})\equiv
\Pi^{(B)}_{ij}({\mathbf k},t_*)a^4_*,
\label{eq:B-turb-Pi-comoving}
\end{equation}
where $\tilde{{\mathbf k}}$ is the comoving wave vector corresponding to the
physical wave vector ${\mathbf k}$ at the time of the phase transition.

\subsection{Gravitational Radiation Power Spectrum}

During the radiation-dominated epoch, $a\propto\eta$ and the homogeneous
solutions to Eq.~(\ref{hij_evolution}) are the zero-order spherical
Bessel functions, i.e.\ $j_0(\tilde{k}\eta)$ and $y_0(\tilde{k}\eta)$.
Defining $x\equiv\tilde{k}\eta$ and $x_*\equiv\tilde{k}\eta_*$,
where $\eta_*$ is the conformal time corresponding to the turbulent
source generating the magnetic field,
the usual Green function technique yields the following inhomogeneous solution
for the radiation-dominated epoch:
\begin{equation}
h^{(B)}_{ij}(\tilde{{\mathbf k}},\eta)
=\frac{8\pi G\Pi^{(B)}_{ij}(\tilde{\mathbf k})}{\tilde{k}^2}\int^x_{x_*}dx'\,
\frac{j_0(x')y_0(x)-y_0(x')j_0(x)}{a^2W(x')},
\label{eq:B-turb-RD-soln-1}
\end{equation}
where $W$ is the Wronskian of the homogeneous solutions
\begin{equation}
W(x)=j_0(x)\frac{d}{dx}y_0(x)-y_0(x)\frac{d}{dx}j_0(x)=\frac{1}{x^2}.
\label{eq:B-turb-Wronskian}
\end{equation}
Note that in the turbulence case, the time dependence of the
turbulent source is known only statistically.  The magnetic field, however,
is a {\it coherent} source, and it evolves by frozen flux
until being damped
away by neutrino viscosity.   Therefore in writing down the
gravitional wave equation inhomogeneous solution in
Eq.~(\ref{eq:B-turb-RD-soln-1}), the explicit time dependence of the magnetic
source is known and we can immediately perform the time integral, unlike
the turbulence case.  Substituting Eq.~(\ref{eq:B-turb-Wronskian})
into Eq.~(\ref{eq:B-turb-RD-soln-1}),
using the explicit expressions for the zero-order spherical Bessel functions,
i.e.\ $j_0(x)=\sin x/x$ and $y_0 = -\cos x/x$, and the approximation for
the scale factor in the radiation-dominated epoch
$a(\eta)\simeq H_0\eta\sqrt{\Omega_{\text{rad}}}$, we obtain
\begin{equation}
h^{(B)}_{ij}(\tilde{\mathbf k},\eta)
\simeq\frac{8\pi G\Pi^{(B)}_{ij}(\tilde{\mathbf k})}
{\tilde{k}\eta H^2_0\Omega_{\text{rad}}}
\xi(\tilde{k},\eta_*,\eta),
\qquad\eta\leq\eta_{\tilde{k}},
\label{eq:B-turb-RD-soln-3}
\end{equation}
where $\eta_{\tilde{k}}$ corresponds to the conformal time at which the
magnetic perturbation comoving wave number $\tilde{k}$ is damped away
by neutrino viscosity.   
Here we have abbreviated
\begin{eqnarray}
\xi(\tilde{k},\eta_*,\eta)=\xi(\tilde{k}\eta_*,\tilde{k}\eta)
\equiv\int^\eta_{\eta_*}d\eta'\,
\frac{\sin[\tilde{k}(\eta-\eta')]}{\eta'}.
\label{eq:B-turb-xi-def}
\end{eqnarray}
It is simple to see that $\xi$ is an oscillating function with
a monotonically decreasing amplitude of oscillation; the amplitude
decays more slowly than the $\eta^{-1}$ dependence of free gravitational
waves, since the wave is continually sourced by the magnetic field. 

As in the turbulence case, we are interested in the
average power spectrum of the waves, so we consider the quantity
\begin{equation}
\langle h^{(B)}_{ij}(\tilde{{\mathbf k}},\eta)
h^{(B)*}_{ij}(\tilde{{\mathbf k}}',\eta)\rangle
\simeq\left[\frac{8\pi G}{\tilde{k}\eta H^2_0\Omega_{\text{rad}}}
\xi(\tilde{k},\eta_*,\eta)\right]^2
\langle \Pi^{(B)}_{ij}(\tilde{\mathbf k})\Pi^{(B)*}_{ij}(\tilde{\mathbf k}')
\rangle.
\label{eq:B-turb-2pt-fcn}
\end{equation}
{}From Eq.~(\ref{eq:B-turb-Pi-comoving}), we have
\begin{equation}
\langle \Pi^{(B)}_{ij}(\tilde{\mathbf k})\Pi^{(B)*}_{lm}(\tilde{\mathbf k}')
\rangle
= a^8_*
\langle \Pi^{(B)}_{ij}({\mathbf k},t_*)\Pi^{(B)*}_{lm}({\mathbf k}',t_*)
\rangle.
\label{eq:B-turb-Pi-2pt-fcn}
\end{equation}
As in Eq.~(2.20) of Ref.~\cite{mack02}, the two-point correlation function
$\langle \Pi^{(B)}_{ij}({\mathbf k},t_*)\Pi^{(B)*}_{lm}({\mathbf k}',t_*)
\rangle$ at the time of the phase transition can be written as:
\begin{equation}
\langle\Pi^{(B)}_{ij}({\mathbf k},t_*)\Pi^{(B)*}_{lm}
({\mathbf k'},t_*)\rangle
\equiv \frac{{\mathcal M}_{ijlm}(\hat{{\mathbf k}})}{V}|\Pi^{(B)}(k,t_*)|^2
\delta({\mathbf k-k'}),
\label{eq:B-turb-2pt-fcn-tensor1} 
\end{equation}
where the tensor structure ${\mathcal M}_{ijlm}$ is
[Eq.~(2.21) of Ref.~\cite{mack02}]
\begin{eqnarray}
{\mathcal M}_{ijlm}(\hat{{\mathbf k}})
&\equiv& P_{il}(\hat{{\mathbf k}})P_{jm}(\hat{{\mathbf k}})+
P_{im}(\hat{{\mathbf k}})P_{jl}(\hat{{\mathbf k}})
-P_{ij}(\hat{{\mathbf k}})P_{lm}(\hat{{\mathbf k}}) \nonumber\\ 
&=& \delta_{il}\delta_{jm}+\delta_{im}\delta_{jl}
-\delta_{ij}\delta_{lm}
+{\hat k}_i{\hat k}_j{\hat k}_l{\hat k}_m \nonumber\\
& & \mbox{}+\delta_{ij}{\hat k}_l{\hat k}_m+\delta_{lm}{\hat k}_i{\hat k}_j
-\delta_{il}{\hat k}_j{\hat k}_m-\delta_{jm}{\hat k}_i{\hat k}_l
-\delta_{im}{\hat k}_j{\hat k}_l-\delta_{jl}{\hat k}_i{\hat k}_m
\label{eq:B-turb-M_ijlm}
\end{eqnarray}
and satisfies ${\mathcal M}_{ijij}=4$ and 
${\mathcal M}_{iilm} = {\mathcal M}_{ijll} = 0$.
Then using Eqs.~(\ref{eq:B-turb-stress}), (\ref{Piij}), and (\ref{Pij}),
a similar calculation
as in the previous section (see also the Appendix of Ref.~\cite{mack02}) gives
\begin{equation}
\langle\Pi^{(B)}_{ij}({\mathbf k},t_*)
\Pi^{(B)*}_{ij}({\mathbf k'},t_*)\rangle
=\frac{1}{(4\pi)^2}\delta({\mathbf k-k'})\int d{\mathbf q}\,
P_B(q)P_B(|{\mathbf k-q}|)(1+\gamma^2)(1+\beta^2),
\label{eq:B-turb-Pi-2pt-fcn-PT}
\end{equation}
where as in Eq.~(\ref{pipi2}) we have defined $\gamma=\hat{{\mathbf k}}\cdot
\hat{{\mathbf q}}$ and $\beta=\hat{{\mathbf k}}\cdot\widehat{{\mathbf k-q}}$.
In deriving Eq.~(\ref{eq:B-turb-Pi-2pt-fcn-PT}), we have assumed the
turbulence-induced magnetic field to be Gaussian, as in the case of the
turbulent fluid velocities, and hence we can apply Wick's theorem.  Comparing
with Eq.~(\ref{eq:B-turb-2pt-fcn-tensor1}), replacing $\gamma^2$
and $\beta^2$ by their average values over the integral of $1/2$, 
and using Eq.~(\ref{eq:B-turb-power-spectrum}) gives
\begin{equation}
|\Pi^{(B)}(k,t_*)|^2\simeq\frac{9V}{16}w^2
\int d{\mathbf q}\,P(q)P(|{\mathbf k-q}|).
\label{eq:B-turb-Pi-corr-spectrum2}
\end{equation}
This integral has already been done in Eq.~(\ref{PPintegral}),
except that now the lower limit for the physical wave number
is $k_B$ instead of $k_S$; hence
\begin{equation}
|\Pi^{(B)}(k,t_*)|^2\simeq\frac{27\pi^5}{8V}w^2
\bar{\varepsilon}^{4/3}k^{-11/3}k^{-2/3}_B.
\label{eq:B-turb-Pi-corr-spectrum3}
\end{equation}

Equations (\ref{eq:B-turb-2pt-fcn}), (\ref{eq:B-turb-Pi-2pt-fcn}),
(\ref{eq:B-turb-2pt-fcn-tensor1}), and (\ref{eq:B-turb-Pi-corr-spectrum3})
together then give 
\begin{equation}
\langle h^{(B)}_{ij}(\tilde{{\mathbf k}},\eta)
h^{(B)*}_{ij}(\tilde{{\mathbf k}}',\eta)\rangle
\simeq \frac{864\pi^7G^2}{\tilde{V}^2}
\frac{w^2\bar{\varepsilon}^{4/3}\tilde{k}^{-17/3}\tilde{k}^{-2/3}_B
a^{28/3}_*}{a^2H^2_0\Omega_{\text{rad}}}
\xi^2({\tilde k},\eta_*,\eta)
\delta(\tilde{{\mathbf k}}-\tilde{{\mathbf k}}'),
\label{eq:B-turb-2pt-fcn-soln}
\end{equation}
where we have approximated $a\simeq H_0\eta\sqrt{\Omega_{\text{rad}}}$
while the Universe is radiation dominated, and we have converted to
the comoving quantities $\tilde{V}=V/a^3_*$, $\tilde{k}=ka_*$,
$\delta(\tilde{\mathbf k}-\tilde{\mathbf k}')=
\delta({\mathbf k}-{\mathbf k}') /a^3_*$.
As in the previous section, we evaluate the real-space
correlation function to
make contact with measurable quantities:
\begin{eqnarray}
\langle h^{(B)}_{ij}(\tilde{{\mathbf x}},\eta)
h^{(B)}_{ij}(\tilde{{\mathbf x}},\eta)\rangle
&=     & \frac{\tilde{V}^2}{(2\pi)^6}\int d\tilde{{\mathbf k}}
d\tilde{{\mathbf k}}'\,e^{i(\tilde{{\mathbf k}}'-\tilde{{\mathbf k}})
\cdot\tilde{{\mathbf x}}} \langle h^{(B)}_{ij}(\tilde{{\mathbf k}},\eta)
h^{(B)*}_{ij}(\tilde{{\mathbf k}}',\eta)\rangle \nonumber\\
&\simeq& \frac{54\pi^2 G^2w^2\bar{\varepsilon}^{4/3}
\tilde{k}^{-2/3}_Ba^{28/3}_*}{a^2H^2_0\Omega_{\text{rad}}}
\int^{\tilde{k}_D}_{\tilde{k}_B}d\tilde{k}\,\tilde{k}^{-11/3}
\xi^2({\tilde k},\eta_*,\eta)
\label{eq:B-turb-2pt-fcn-real-1}
\end{eqnarray}

\subsection{Relic Gravitational Radiation}

As in Eq.~(\ref{hcdef}), we define the characteristic gravitational wave
amplitude $h^{(B)}_c(\tilde{f})$ per unit logarithmic comoving
frequency interval via
\begin{equation}
\langle h^{(B)}_{ij}(\tilde{{\mathbf x}},\eta)
h^{(B)}_{ij}(\tilde{{\mathbf x}},\eta)\rangle
\equiv \frac{1}{2}\int^\infty_0\frac{d\tilde{f}}{\tilde{f}} 
h^{(B)2}_c(\tilde{f},\eta),
\label{eq:B-turb-char-amplitude-def}
\end{equation}
and the statistical average on the left side implies that the
time dependence on the right side is not the exact time dependence of
the gravitational wave but only the time dependence of its amplitude
(i.e., the oscillations are averaged over).
Since Eq.~(\ref{eq:B-turb-RD-soln-3}) 
gives the exact time dependence of the gravitational
radiation (as opposed to the turbulence case, when we only know
the statistically averaged time dependence), the standard wave dispersion
relation holds: ${\tilde k} = 2\pi {\tilde f}$. (In contrast, 
for the turbulence
source, we only know the statistically averaged time dependence, so we
use only the approximate dispersion relation $f = \tau_L^{-1}$.)
Then comparing with Eq.~(\ref{eq:B-turb-2pt-fcn-real-1}) gives
\begin{equation}
h^{(B)}_c(\tilde{f},\eta_0) \simeq 
\left(\frac{3^{3/2}}{2^{2/3}\pi^{2/3}}\right)
\frac{Gw\bar{\varepsilon}^{2/3}\tilde{f}^{-1/3}_B a^{14/3}_*}
{H_0\sqrt{\Omega_{\text{rad}}}}\tilde{f}^{-4/3} 
\bar{\xi}(\tilde{f}\eta_*,\tilde{f}\eta_{\rm end}),
\label{eq:B-turb-char-amplitude-1}
\end{equation}
which depends on the function $\bar{\xi}$ which we define as
the amplitude of the oscillations in $\xi(\eta)$, times the
numerical factor $\sqrt{2}/2$ to convert to a root-mean-square value,
in accordance with the definition of the characteristic amplitude
Eq.~(\ref{eq:B-turb-char-amplitude-def}).
In deriving Eq.~(\ref{eq:B-turb-char-amplitude-1}), 
we have used the fact that after the conformal time $\eta_{\rm end}$
when the magnetic fields are damped away via viscosity and cease to be
an efficient source of gravitational radiation, the characteristic
amplitude $h^{(B)}_c(\tilde{f},\eta)$ will simply scale inversely with $a$.
Note that $h^{(B)}_c(\tilde{f},\eta_0)$ is only weakly dependent on
$\eta_{\text{end}}$, which occurs through the upper limit of the integral in
$\xi(\tilde{f},\eta_*,\eta_{\text{end}})$. The time $\eta_{\rm end}$
technically depends on the scale considered, but for simplicity
we simply use the saturation scale $L_B$ at which the gravitational
radiation peaks. 

The function $\bar\xi({\tilde k}\eta_*,{\tilde k}\eta)$ 
cannot be expressed in terms
of elementary functions, but it is simple to obtain an
upper bound. Since the amplitude of the oscillations in $\xi$ 
is monotonically decreasing, the amplitude at the initial time
gives 
\begin{equation}
\bar\xi({\tilde f}\eta_*,{\tilde f}\eta) < 
{\sqrt{2}\over 4\pi{\tilde f}\eta_*},
\label{xibound}
\end{equation}
which is useful for constraining the gravitational wave amplitude.

In an analogous calculation to the previous section, writing
$\tilde{f}_B=f_Ba_*$ and $f_B=L^{-1}_B$ and approximating
$H_*\simeq H_0 \sqrt{\Omega_{\text{rad}}}a^{-2}_*$,
we obtain 
\begin{equation}
h^{(B)}_c(\tilde{f},\eta_0)
\simeq 1.9\times10^{-16}\left(\frac{\kappa\rho_{\text{vac}}}{w}\right)^{2/3}
\left(\frac{\tau}{H^{-1}_*}\right)^{-2/3}
\left(\frac{L_B}{H^{-1}_*}\right)^{5/3}
\left(\frac{100\,\text{GeV}}{T_*}\right)
\left(\frac{100}{g_*}\right)^{1/3}
\left(\frac{\tilde{f}}{\tilde{f}_B}\right)^{-4/3}
\bar{\xi}(\tilde{f},\eta_*,\eta_{\text{end}}).
\label{eq:B-turb-char-amplitude-3}
\end{equation}
This characteristic amplitude is valid for $\tilde{f}>\tilde{f}_B$, where
[using Eq.~(\ref{tildef})]
\begin{equation}
\tilde{f}_B=1.65\times10^{-5}\,\text{Hz}
\left(\frac{L_B}{H^{-1}_*}\right)^{-1}
\left(\frac{T_*}{100\,\text{GeV}}\right)
\left(\frac{g_*}{100}\right)^{1/6}
\label{eq:B-turb-char-frequency}
\end{equation}
The corresponding
energy density in gravitational waves per logarithmic frequency
interval in units of current critical density is
\begin{equation}
\Omega^{(B)}_{\text{GW}}(\tilde{f})h^2
=6.0\times10^{-6} \left(\frac{\kappa\rho_{\text{vac}}}{w}\right)^{4/3}
\left(\frac{\tau}{H^{-1}_*}\right)^{-4/3}
\left(\frac{L_B}{H^{-1}_*}\right)^{4/3}
\left(\frac{g_*}{100}\right)^{-1/3}
\left(\frac{\tilde{f}}{\tilde{f}_B}\right)^{-2/3}
\bar{\xi}^2(\tilde{f},\eta_*,\eta_{\text{end}}).
\label{eq:B-turb-char-energy}
\end{equation}

\section{First-Order Cosmological Phase Transitions}
\label{sec:pt}

The most likely mechanism for creating turbulence with a large energy
density is a first-order phase transition.  Such a transition is
controlled by an effective potential for some quantity which functions
as the order parameter of the phase transition. Initially, the
Universe sits in a minimum of the effective potential.  As the
Universe expands and cools, the effective potential develops a local
minimum at a different value of the order parameter; this new local
minimum eventually evolves to be the true minimum energy state.  Then
the order parameter wants to evolve to the new minimum. If a potential
energy barrier exists between the old local minimum and the new true
minimum, the phase transition must occur via quantum tunnelling
through the barrier or thermal fluctuations over the barrier. As a
result, bubbles of the low-temperature phase are nucleated at random
places in the high-temperature phase. The energy difference between
the two phases creates an effective outward force on the bubble,
causing it to expand. Once this outward force from the energy
difference balances the inward hydrodynamic force from pushing plasma
outwards, the bubble reaches an equilibrium and expands at a constant
velocity.  We will consider only the case of quantum tunnelling,
applicable to a strong first-order phase transition with a high
barrier between the two phases. In this case the nucleated bubbles are
spherical and negligibly small compared to the horizon scale
\cite{col77}. The more complex case of thermally activated bubbles 
has been considered in \cite{ghk97}.

\subsection{Turbulence}

In general, the rate for nucleating a bubble will be the exponential
of some tunnelling action, $\Gamma\propto \exp(S(t))$.  As a simple
model of a phase transition, we expand the action $S$ into a power
series in time and keep only the constant and linear terms. This gives
a characteristic bubble nucleation rate per unit volume \cite{tww92}
\begin{equation}
\Gamma = \Gamma_0 e^{\beta t}
\label{gamma_def}
\end{equation}
so the quantity $\beta^{-1}$ sets the characteristic time scale for
the phase transition. Numerical calculations show that the largest
bubbles reach a size of order $\beta^{-1} v_b$ by the end of the phase
transition \cite{kt93}, where $v_b$ is the bubble expansion velocity,
assuming the bubbles remain spherical as they expand. In
general, $\beta$ is expected to be of the order $4\ln(m_{\rm Pl}/T) H
\simeq 100 H$ for a Hubble rate $H$ \cite{tww92}.

A first-order phase transition is generically described by several
parameters: (i) $\alpha \equiv \rho_{\rm vac} / \rho_{\rm thermal} 
= 4\rho_{\rm vac} / 3w$, the ratio of the
vacuum energy associated with the phase transition to the thermal
density of the Universe at the time (which characterizes the strength
of the phase transition); (ii) $\kappa$, an efficiency
factor which gives the fraction of the available vacuum energy which
goes into the kinetic energy of the expanding bubble walls, as opposed
to thermal energy; (iii) $\beta$, which sets the characteristic time
scale for the phase transition; (iv) $v_b$, the velocity of the
expanding bubble walls, which set the characteristic length scale of
the phase transition; (v) $T_*$, the temperature at which the phase
transition occurs.

Once the bubbles expand and percolate, much of their kinetic energy
will be converted to turbulent bulk motions of the primordial plasma
(for an illustration, see the numerical evolution of two scalar field
bubbles in Ref.~\cite{ktw92a}). The energy density contained in a
bubble wall of radius $r$ scales with $r^3$, the bubble volume. As the
phase transition ends, far more small bubbles have been nucleated than
large ones, but the energy density in the large ones dominates the
total energy density \cite{tww92}. We therefore make the approximation
that turbulent energy is injected on a stirring scale $L_S\simeq
v_b\beta^{-1}$ corresponding to the size of the largest bubbles. The
stirring will last for roughly $\tau_{\rm stir}=\beta^{-1}$, the
duration of the phase transition. The duration $\tau$ of the
turbulence then follows from Eqs.~(\ref{taudef}) and (\ref{kDvalue2}) as
\begin{equation}
\tau = \beta^{-1}\,{\rm max}\left[1,\,
{3\sqrt{2}\over 2} {v_b\over(\kappa\alpha)^{1/2}}\right].
\label{tau_turb}
\end{equation}

The fundamental symmetry breaking mechanism which drives the phase
transition determines some effective potential for bubble
nucleation. The difference in energy density between the two phases
and the bubble nucleation rate are both determined by this mechanism.
Thus the parameters $T_*$, $\beta$, and $\alpha$ are all determined
directly by the underlying physics, and are precisely calculable to
some given order in the various particle interaction strengths. On the
other hand, the bubble velocity $v_b$ and the fraction of kinetic
energy into the bubbles $\kappa$ depend on the detailed microphysics
involved in the bubble propagation through the relativistic plasma and
are not determined from general properties of the effective potential.
Generally, the larger the vacuum energy density driving phase
transition, the higher bubble wall velocities $v_b$ will be obtained.

The hydrodynamic boundary between a lower-energy phase and a
higher-energy one can propagate via two modes, detonation and
deflagration. Details of these modes in the case of spherical geometry
are known \cite{ste82}. For a detonation front, the velocity of the
phase boundary exceeds the sound speed in the fluid, so that a shock
forms at the burning front. In the opposite case, a deflagration
propagates slower than the sound speed and piles up an overdensity of
fluid in front of it, like a snowplow. The boundary conditions for a
detonation are more restrictive, so that once the energy densities and
pressures are specified in each phase, the complete solution for the
propagating detonation is determined. In this case, we have
\cite{ste82}
\begin{equation}
v_b(\alpha) = {1/\sqrt{3} + (\alpha^2 + 2\alpha/3)^{1/2} \over 1+\alpha}
\label{vb_alpha}
\end{equation}
and the approximate form \cite{kkt94}
\begin{equation}
\kappa(\alpha) = {1\over 1+A\alpha}
\left[A\alpha + {4\over 27}\left(3\alpha\over 2\right)^{1/2}\right]
\label{kappa_alpha}
\end{equation}
with $A=0.72$. If the bubbles propagate as a deflagration front, no
such general relations apply. However, it has been argued that for
relativistic plasmas, instabilities in the bubble shape will
accelerate the bubble walls and the hydrodynamic expansion mode is
unstable to becoming a detonation. For this reason, in the following
analysis, we will assume Eqs.~(\ref{vb_alpha}) and (\ref{kappa_alpha})
hold. We also assume that $\alpha \ll 1$ to simplify further
Eqs.~(\ref{vb_alpha}) and (\ref{kappa_alpha}), which will generally
hold for realistic phase transition models; for unusual cases with
very strong detonations and $\alpha\gtrsim 1$, the following formulas
must be corrected. The duration of the turbulence is then given by the
second term in Eq.~(\ref{tau_turb}), becoming
\begin{equation}
\tau = \left(3\over 2\right)^{9/4}\beta^{-1}\alpha^{-3/4}.
\label{tau_det}
\end{equation}

\subsection{Relic Radiation from the Phase Transition}

The characteristic gravitational wave amplitude from turbulence becomes
\begin{equation}
h_c({\tilde f}) \simeq 3.8\times 10^{-18} \alpha^{9/8}
\left(H_*\over\beta \right)^2
\left(100\,{\rm GeV}\over T_*\right)
\left(100\over g_*\right)^{1/3}
\left({\tilde f}\over {\tilde f}_S\right)^{-11/4},
\label{hc_det}
\end{equation}
with the characteristic frequency
\begin{equation}
{\tilde f}_S \simeq 5.7\times 10^{-6}\,{\rm Hz}\, \alpha^{3/4}
\left(\beta\over H_*\right)\left(T_*\over 100\,{\rm GeV}\right)
\left(g_*\over 100\right)^{1/6}.
\label{fs_det}
\end{equation}
The corresponding energy density per logarithmic frequency interval
is
\begin{equation}
\Omega_{\rm GW} h^2 \simeq 2.7\times 10^{-10} \alpha^{15/4}
\left(H_*\over\beta\right)^2
\left(g_*\over 100\right)^{-1/3}
\left({\tilde f}\over {\tilde f}_S\right)^{-7/2}.
\label{omega_det}
\end{equation}

In a first-order phase transition, the expanding, colliding bubbles
are themselves a potent source of gravitational radiation
\cite{kkt94}. For our idealized model phase transition with spherical
expanding bubbles, the ratio of the maximum amplitude of gravitational
radiation due to turbulence to the maximum amplitude due to bubble
collisions is approximately
\begin{equation}
{h_{\rm turb}({\tilde f}_S)\over 
h_{\rm bub}({\tilde f}_{\rm max})} \simeq 0.18 \alpha^{-3/8},
\label{hratio}
\end{equation}
so only for $\alpha < 0.01$ will the amplitude of the
turbulent signal be larger (although in this case, turbulent damping
due to the expansion of the Universe is significant and our estimate
for the turbulence gravitational wave amplitude may be significantly
too large).  For realistic models with interesting gravitational wave
production, the turbulence amplitude will be subdominant to the bubble
amplitude, but non-negligible.  This is in contrast to the naive
dimensional estimate of the turbulence gravitational radiation in
Ref.~\cite{kkt94} which gave a somewhat larger value. The frequencies
at which these maximum amplitudes occur scale differently with the
parameters:
\begin{equation}
{{\tilde f}_S\over{\tilde f}_{\rm max}} = 1.1 \alpha^{3/4}.
\label{fratio}
\end{equation}
The different scaling arises because the duration of the phase
transition $\tau_{\rm stir}$ sets the characteristic frequency for the
radiation from expanding bubbles, while the circulation time on the
stirring scale $\tau_S$ sets the characteristic frequency for the
radiation from turbulence.

Note that the gravitational radiation in the bubble case has a long
tail in the amplitude, $h_c(f)\propto f^{-1/3}$, while turbulence
driven at a single scale drops off very quickly like $h_c(f)\propto
f^{-11/4}$.  The tail for bubble collisions arises in the case of
bubble collisions because at any given moment, the characteristic
frequency of radiation from the collision of two bubbles is $v_b/d$,
where $d$ is the size of the colliding region. Since $d$ ranges from
zero to the maximum size of the smaller bubble as the bubbles expand,
the gravitational radiation is produced over a wide range of
frequencies. This tail of the frequency spectrum is somewhat
model-dependent, and will be modified if the bubbles are not
spherical. Departures from sphericity could arise from thermal
activation over the potential barrier, resulting in non-spherical
nucleation, or from shape instabilities as the bubble expands. The
results for expanding bubbles also depend on the thin-wall
approximation, namely that the width of the bubble wall is small
compared to the radius of the bubble. While this approximation will be
very good for relativistic detonations, it will not be as good for
deflagrations.

The gravitational wave signal from turbulence from a single stirring
scale $L_S$ is somewhat more generic, although if the phase transition
does not proceed via detonation the specific expressions for $\kappa$
and $v_b$ in Eqs.~(\ref{kappa_alpha}) and (\ref{vb_alpha}) will not
hold. However, the single-scale assumption obviously will never be
exactly correct; any realistic source like a phase transition will
deposit bulk kinetic energy over a range of scales. The energy density
in bubble walls of a given size will generically peak at a scale
comparable to $v_b \beta^{-1}$ that we have taken for $L_S$, because
the kinetic energy in the wall of a bubble of radius $r$ scales like
$r^3$ so the energy distribution is heavily weighted towards the
largest bubbles. Analytic expressions for the size distribution of
bubbles, the fraction of space taken up by bubbles, and related
quantities are given in Ref.~\cite{tww92}.  On the other hand, the
stirring scale appropriate to the collision of two bubbles of unequal
radius is not entirely clear: some turbulence will clearly be created
on the scale of the smaller bubble, but since the larger bubble has
greater energy density in the wall, a significant part of the energy
will remain in coherent motion determined by the larger bubble.

In realistic cases, the gravitational wave amplitude spectrum in
Eq.~(\ref{hc_det}) must be convolved over a range of stirring
scales. A specific model of the distribution of stirring scales in a
first-order phase transition is beyond the scope of this
paper. However, we can make a rough estimate of its effect.  Assume
that the actual turbulence source stirs the plasma over a range of
frequencies $\Delta f_S$. The actual bubble size distribution has a
significant tail towards larger bubbles \cite{tww92}.  If the same
total energy goes into gravitational radiation as in the single
stirring scale case, then the characteristic amplitude $h_c(f_S)$ will
be reduced by a factor of order $(f_S/\Delta f_S)^{1/2}$.  This very
crude estimate neglects the strong dependence of the amplitude on the
stirring scale and employs only a box-shaped energy density spectrum,
but the general scaling is correct. Generically, the distribution of
bubble sizes in a model phase transition points to $\Delta f_S /f_S$ 
on the order of a few (see \cite{tww92}), but a more precise estimate
requires a detailed model of stirring in a phase transition. As a rule
of thumb, when estimating the gravitational radiation background from
turbulence arising from a phase transition with a single
stirring-scale model of the turbulence, the resulting amplitude may be
overestimated by a modest factor.

\subsection{Relic Radiation from the Induced Magnetic Fields}

For the magnetic fields from the turbulent dynamo mechanism,
the characteristic gravitational wave amplitude
becomes 
\begin{equation}
h^{(B)}_c(\tilde{f},\eta_0) \simeq
1.0\times10^{-17} \alpha^{3/2}
\left(\frac{H_*}{\beta}\right) \left(\frac{L_B}{L_S}\right)^{5/3}
\left(\frac{100\,\text{GeV}}{T_*}\right)
\left(\frac{100}{g_*}\right)^{1/3}
\left(\frac{\tilde{f}}{\tilde{f}_B}\right)^{-4/3}
\bar{\xi}(\tilde{f}\eta_*,\tilde{f}\eta_{\text{end}}),
\label{eq:B-turb-char-amplitude-1st}
\end{equation}
with the characteristic frequency
\begin{equation}
\tilde{f}_B \simeq
2.9\times10^{-5}\,\text{Hz}
\left(\frac{L_S}{L_B}\right)
\left(\frac{\beta}{H_*}\right)
\left(\frac{T_*}{100\,\text{GeV}}\right)
\left(\frac{g_*}{100}\right)^{1/6}.
\label{eq:B-turb-char-frequency-1st}
\end{equation}
The corresponding energy density per logarithmic frequency interval is
\begin{equation}
\Omega^{(B)}_{\text{GW}}(\tilde{f})h^2 \simeq
4.8\times10^{-8} \alpha^3
\left(\frac{L_B}{L_S}\right)^{4/3}
\left(\frac{g_*}{100}\right)^{-1/3}
\left(\frac{\tilde{f}}{\tilde{f}_B}\right)^{-2/3}
\bar{\xi}^2(\tilde{f}\eta_*,\tilde{f}\eta_{\text{end}}).
\label{eq:B-turb-char-energy-1st}
\end{equation}

The ratio of the maximum amplitude of gravitational radiation due to
turbulence-induced magnetic fields to the maximum amplitude due to the
turbulent fluid today is approximately
\begin{equation}
\frac{h^{(B)}_c(\tilde{f}_B)}{h^{(\text{turb})}_c(\tilde{f}_S)}
\simeq 2.7 \alpha^{3/8} \left(\frac{\beta}{H_*}\right)
\left(\frac{L_B}{L_S}\right)^{5/3} 
\bar{\xi}(\tilde{f}_B\eta_*,\tilde{f}_B\eta_{\text{end}}).
\label{eq:B-turb-amplitude-ratio} 
\end{equation}
The ratio of the frequencies at which these maximum amplitudes occur is
\begin{equation}
\frac{\tilde{f}_B}{\tilde{f}_S}
\simeq 5.1 \alpha^{-3/4} \left(\frac{L_S}{L_B}\right).
\label{eq:B-turb-frequency-ratio}
\end{equation}
The scaling with $\alpha$ arises because
the circulation time on the stirring scale $\tau_S$
sets the characteristic frequency for the radiation from turbulence, whereas
for magnetic fields, $f_B=L^{-1}_B$.  

The value of $\eta_{\rm end}$ corresponding to the scale $L_B$ can
be determined via consideration of the neutrino viscosity
(see \cite{cd02}) but 
$\bar\xi({\tilde k}\eta_*,{\tilde k}\eta_{\rm end})$ is only
weakly dependent on $\eta_{\rm end}$ so we do not compute it here.
Instead, we derive an upper bound on the amplitude. The approximate
relation $a_* H_* \simeq 1/\eta_*$, valid during radiation domination,
gives ${\tilde f}_B \eta_* \simeq H_*^{-1} / L_B$. Then Eq.~(\ref{xibound})
combined with the frequency dependence in Eq.~(\ref{hc_det}) gives
\begin{equation}
\frac{h^{(B)}_c(\tilde{f}_B)}{h^{(\text{turb})}_c(\tilde{f}_B)}
< 27 \alpha^{-27/16} v_b \left({L_S\over L_B}\right)^{1/12}.
\label{hbound}
\end{equation}
As discussed above, $L_S/L_B\simeq 300$ generically, so
the peak characteristic amplitude from the magnetic field at
frequency ${\tilde f}_B$ will always be negligible compared
to the peak characteristic amplitude from the turbulence at
frequency ${\tilde f}_S$. The turbulence gravitational waves
drop so quickly with frequency, however, that the magnetic
field gravitational waves will give a larger characteristic
amplitude at ${\tilde f}_B$. 

\section{Potential Detectability}
\label{sec:detect}

The detectability of a given stochastic background depends on both its
characteristic frequency and its amplitude.  The Laser Interferometer
Gravitational-wave Observatory (LIGO) \cite{LIGO} is nearing the
commencement of scientific observations; it is comprised of two
facilities in the United States, each essentially a Michelson
interferometer with an arm length of 4 kilometers. LIGO has
sensitivity to gravitational radiation in the frequency range from 10
to 1000 Hz.  Seismic noise prevents useful gravitational wave
detection from the surface of the Earth at frequencies lower than
about 10 Hz.  Cross-correlation of the two LIGO detectors, along with
several smaller laser interferometers and bar detectors at other sites
around the world, allow a clean detection of stochastic signals, since
widely separated detectors have no correlated sources of
noise. Detailed estimates shows that in this frequency range, LIGO
will be able to detect stochastic gravitational wave backgrounds with
a characteristic
amplitude of around $h_c({\tilde f})\simeq 3\times 10^{-23}$ 
at ${\tilde f}\simeq 100$ Hz after integrating for
four months \cite{mic87,chr93,fla93,ar99}. These levels will hopefully be
obtained within three years. Planned technical improvements are
projected to reduce this threshold amplitude by another factor of 10
on the time scale of a decade.

The other major gravitational wave observation program, the Laser
Interferometer Space Antenna (LISA) \cite{LISA}, is a cornerstone
mission of the European Space Agency in partnership with NASA. 
Current design
studies envision three spacecraft arrayed in an equilateral triangle
with an arm length of around $5\times 10^6$ kilometers with laser
interferometry between each of the three pairs of arms; the spacecraft
configuration will trail the Earth's orbit by about $20^\circ$. LISA
will likely be sensitive to a frequency range from around 0.0001 Hz to
0.1 Hz. The detection of stochastic backgrounds with LISA is more
complicated than with LIGO, because any pair of interferometers formed
by LISA's arms share one arm in common, so it is not possible to
cross-correlate two independent interferometers with uncorrelated
noise. It was originally believed that this limited detection of a
stochastic background to the level of the instrument noise power
because there would be no way to distinguish between instrumental
noise and a background signal. This noise level corresponds to a
stochastic background amplitude of around $h_c({\tilde f})=10^{-21}$ at 0.01
Hz.  It has now been realized that if the complete time series data
for positions of 6 independent test masses are recorded, so-called
Sagnac observables can be synthesized which are highly insensitive to
various kinds of noise in the system \cite{arm99}, including one which
is largely independent of low-frequency stochastic gravitational wave
backgrounds, allowing a direct measurement of the system noise
\cite{tin00}. This results in a significant improvement in the ability
of the system to measure stochastic backgrounds \cite{hb01}. For one
year of observation, this kind of analysis could in principle give
sensitivities comparable to two independent Michelson interferometers,
reducing the threshold $h_c({\tilde f})$ by a factor of 
$({\tilde f}t)^{1/4}$ for
observation over a time $t$, or $h_c({\tilde f})\simeq 4\times 10^{-23}$ at
0.01 Hz over one year of observing. Such a sensitivity level depends
on a precise understanding of the system noise properties and
elimination of other correlated noise sources between the various arms
of the detector, which is only partially practicable. Flying and
cross-correlating two independent LISA-like detectors
\cite{ben99,uv01} is still clearly preferable for detecting stochastic
backgrounds.

For stochastic background detection at LISA frequencies, raw
sensitivity is not the only issue. White dwarf binaries in our galaxy
will produce an approximately stochastic gravitational wave background
which probably becomes comparable to the LISA sensitivity limits for
frequencies below about $10^{-3}$ Hz \cite{hbw90}.  Detection of such
a signal will be interesting in its own right, but will effectively
provide a lower limit of around $10^{-4}$ Hz to the stochastic
background signals which are detectable, until gravitational wave
detectors improve to the point of having enough directional
sensitivity to distinguish sources in the galactic plane from sources
distributed isotropically.

The characteristic gravitational wave frequency for turbulence from
known phase transitions is not promising for detection in the near
future.  For the electroweak phase transition at $T_*\simeq 100$ GeV,
Eq.~(\ref{fs_det}) shows that $\alpha$, the ratio of the vacuum 
energy density to the thermal density at the time of the
phase transition, must be of order 0.1 for the
frequency maximum to be as high as $\tilde f_S = 10^{-4}$ Hz, if
$\beta/H_*$ takes its characteristic value of 100.  This frequency is
the lower limit to what LISA might be able to detect. The amplitude at
this frequency for $\alpha = 0.1$ would be $h_c({\tilde f}_S) \simeq 2.8
\times 10^{-23}$, more than two orders of magnitude smaller than a LISA
Sagnac configuration could detect at this frequency. Any push towards
higher frequencies via a shorter phase transition further reduces the
characteristic amplitude, since $h_c({\tilde f})\propto
(H_*/\beta)^2$.  For an extreme case with $\alpha=1$ and
$\beta=1000H_*$, the characteristic frequency is near LISA's maximum
sensitivity, ${\tilde f}_S \simeq 5.7\times 10^{-3}$ Hz, with a
characteristic amplitude $h_c({\tilde f}_S) \simeq 3.8\times
10^{-24}$. This amplitude is an order of magnitude smaller
than the LISA Sagnac sensitivity at this frequency. An analysis of the
electroweak effective potential in a large class of supersymmetric
models \cite{pie93,hs01} 
shows that for models with large values of $\alpha$, generally
$\beta < 100 H_*$, and $\alpha$ is never as large as unity
\cite{apr01}. Other well-motivated extensions of the standard
model may result in a very strong electroweak phase transition
(e.g., \cite{acg01}).

Satellite missions to probe stochastic backgrounds to lower
frequencies have been discussed \cite{cl01}, which would involve
multiple spacecraft arrayed at separations on the order of 1 a.u.
Such configurations would be a more natural match for the frequency
scale of electroweak turbulence, although dealing with the binary
foreground signal would still be a major hurdle.

Phase transitions at lower temperatures, like the QCD phase
transitions, have larger characteristic length scales and thus even
lower frequencies for gravitational radiation. Speculative phase
transitions could occur at energy scales higher than the electroweak
scale, resulting in higher characteristic frequencies.  However, a
higher energy scale also translates into a smaller characteristic
amplitude, and it is not possible to give a set of parameters with
$\alpha \lesssim 1$ for which cosmological turbulence would be
detectable in LIGO. LISA could detect the turbulence from a range of
imagined phase transitions at energy scales above the weak scale, but
at present no compelling theoretical motivation for such phase
transitions is at hand.

In contrast to turbulent sources, the expanding bubbles in a
first-order phase transition, which drive the turbulence, are
themselves a strong source of gravitational radiation \cite{kkt94} and
are a much more promising source of detectable signals from the
electroweak phase transition.  The difference between the
detectability of the two sources is essentially the factor of
$\alpha^{3/4}$ in Eq.~(\ref{fratio}), arising from the different time
scales of the sources. The characteristic frequency for the expanding
bubbles is set by the phase transition time scale $\beta^{-1}$ because
the bubbles expand and percolate in this time. For turbulence, the
time scale is instead the circulation time on the stirring scale.  The
turbulent fluid velocities are significantly smaller than the bubble
expansion velocities unless the turbulent flows are near the sound
speed, giving lower characteristic frequencies. (Extremely strong
turbulence with relativistic fluid velocities would likely produce
gravitational radiation in a more detectable range of frequencies and
amplitudes, but the amount of energy in turbulent motions is limited
by shock formation and heating, and turbulence is not understood in
this regime.)  

Our results in this work supercede the dimensional estimates in
Ref.~\cite{kkt94}, which predicted that the turbulence signal could be
significantly larger than the bubble signal at similar
frequencies. Resulting optimistic calculations of turbulent signals
from the electroweak phase transition, e.g.  Ref.~\cite{apr01},
unfortunately do not hold up to more detailed analysis. We emphasize,
however, that the results presented here apply in a generic way to any
turbulence in the early Universe, and the search for stochastic
gravitational radiation backgrounds in the frequency range from
$10^{-4}$ Hz to 1000 Hz is in part a search for unanticipated,
dramatic physics at energies above the electroweak scale.
Perhaps we will be lucky.

\acknowledgments We would like to thank A.~Smilga for discussions
about turbulence, V.~Oudovenko and G.~Chitov for help with numerical
calculations supporting analytical estimates, 
Ruth Durrer for discussions about magnetic fields, Sterl Phinney
for guidance about LISA sensitivities, and Dario Grasso for finding an
errant factor of 2.  John Barrow kindly directed
our attention to numerous historical papers about cosmological
turbulence.  This work has been supported by the NASA Astrophysics
Theory Program through grant NAG5-7015. T.K. has been supported in
part by a grant from the Collaboration in Basic Science and
Engineering Program of the National Research Council.  A.K. is a
Cotrell Scholar of the Research Corporation.

\end{document}